\documentclass{aa}  
\usepackage{graphicx}
\usepackage{amsmath}
\usepackage{txfonts}
\usepackage{natbib}
\bibpunct{(}{)}{;}{a}{}{,} 

\begin{document}

\title{GaBoDS: The Garching-Bonn Deep Survey } \subtitle{VIII.
  Lyman-break galaxies in the ESO Deep Public Survey\thanks{Based on
    observations made with ESO Telescopes at the La Silla
    Observatory.}}  \titlerunning{Lyman-break galaxies in the DPS}


   \author{H. Hildebrandt\inst{1}
          \and
          J. Pielorz\inst{1}
          \and
          T. Erben\inst{1}
          \and
          P. Schneider\inst{1}
          \and
          T. Eifler\inst{1}
          \and
          P. Simon\inst{1}
          \and
          J.\,P. Dietrich\inst{1}
          }

   \offprints{H. Hildebrandt}

   \institute{Argelander-Institut f\"ur Astronomie\thanks{Founded by
       merging of the Sternwarte, Radioastronomisches Institut and
       Institut f\"ur Astrophysik
       und Extraterrestrische Forschung der Universit\"at Bonn}, Universit\"at Bonn, Auf dem H\"ugel 71, D-53121 Bonn, Germany\\
     \email{hendrik@astro.uni-bonn.de} }

   \date{Received ; accepted }

 
  \abstract
   {}
   {The clustering properties of a large sample of $U$-dropouts are
     investigated and compared to very precise results for
     $B$-dropouts from other studies to identify a possible evolution
     from $z=4$ to $z=3$.}
   {A population of $\sim8800$ candidates for star-forming galaxies at
     $z=3$ is selected via the well-known Lyman-break technique from a
     large optical multicolour survey (the ESO Deep Public Survey).
     The selection efficiency, contamination rate, and redshift
     distribution of this population are investigated by means of
     extensive simulations. Photometric redshifts are estimated for
     every Lyman-break galaxy (LBG) candidate from its $UBVRI$
     photometry yielding an empirical redshift distribution.  The
     measured angular correlation function is deprojected and the
     resulting spatial correlation lengths and slopes of the
     correlation function of different subsamples are compared to
     previous studies.}
   {By fitting a simple power law to the correlation function we do
     not see an evolution in the correlation length and the slope from
     other studies at $z=4$ to our study at $z=3$. In
     particular, the dependence of the slope on UV-luminosity similar
     to that recently detected for a sample of $B$-dropouts is
     confirmed also for our $U$-dropouts. For the first time number
     statistics for $U$-dropouts are sufficient to clearly detect a
     departure from a pure power law on small scales down to
     $\sim2\arcsec$ reported by other groups for $B$-dropouts.}
   {}


   \maketitle
%

\section{Introduction}
\label{sec:intro}

For more than a decade now, the high-redshift universe has become
reachable by observations mainly due to the development of efficient
colour selection techniques.  The group around C.\,C.  Steidel
\citep{1993AJ....105.2017S,1996ApJ...462L..17S,1999ApJ...519....1S}
has introduced the Lyman-break technique selecting high-redshift,
star-forming galaxies from optical multicolour data by their
pronounced Lyman-break. Many groups have used this technique and
analysed various properties of these galaxy populations from redshifts
$z\approx3$ up to $z\approx7$.

A particular emphasis in these studies was given to the clustering
properties of the Lyman-break galaxy (LBG) samples
\citep{1998ApJ...492..428S,1998ApJ...503..543G,1998ApJ...505...18A,2005ApJ...619..697A,2001ApJ...550..177G,2001ApJ...558L..83O,2004ApJ...611..685O,2005ApJ...635L.117O,2002ApJ...565...24P,2004ApJ...609..513B,2003A&A...409..835F,2005MNRAS.360.1244A,2006ApJ...642...63L,2006ApJ...637..631K,2006ApJS..162....1G}.
Going back in cosmic time the correlation strength of these high-$z$
galaxies can be directly compared to $N$-body simulations or
semi-analytical predictions yielding estimates of, e.g., the galaxy
bias for early epochs. These results can then be used as an input for
models of galaxy formation, and help to constrain the large number of
free parameters to adjust. The more precise our knowledge of the
clustering evolution becomes the more accurate our understanding of
galaxy evolution will be.

In this context it is of crucial importance to reach a similar
precision for the same galaxy populations at different redshifts.
While in the beginning most LBG studies concentrated on relatively
bright $z=3$ $U$-dropouts, in the last years more and more groups have
focused on the investigation of LBGs at redshifts around $z=4$
selected as $B$-dropouts. This is obviously due to the fact that deep
and wide $U$-band images are still a very telescope-time consuming
task and some recent wide-field cameras like \emph{Suprimecam} or
space-based surveys like GOODS even lack a $U$-filter entirely.

By estimating the angular correlation function of nearly 17\,000
$B$-dropouts selected from the Subaru/XMM-Newton Deep Field,
\cite{2005ApJ...635L.117O} find evidence for a departure from a pure
power law on small scales. The same trend is reported by
\cite{2006ApJ...642...63L} for $B$- and $V$-dropouts from the very
deep GOODS ACS data. Neither the number statistics in the former study
mentioned nor the depth and angular resolution of the GOODS data have
a comparable counterpart at slightly lower redshifts. The most precise
estimates of $U$-dropout clustering to date come from
\cite{2005ApJ...619..697A} estimating the angular correlation function
but not reporting any obvious excess on small scales. The question
whether this is an evolutionary effect or whether this feature is only
visible in the $z=4$ data because of their superior quality can only
be answered by a $U$-dropout survey comparable in size to the
$B$-dropout surveys mentioned above.

In this paper we describe our investigations of $z=3$ LBGs in the
ESO Deep Public Survey (DPS). The methods presented here are based on
our investigations in the Chandra Deep Field South (the DPS field
Deep2c) presented in \cite{2005A&A....Hildebrandt}. In
Sect.~\ref{sec:data} the data of the DPS and the selection of the LBGs
are described. Simulations to assess the performance of our LBG
selection are presented in Sect.~\ref{sec:simulations}. The clustering
ana\-ly\-sis is covered in Sect.~\ref{sec:clustering}.  A summary and
conclusions are given in Sect.~\ref{sec:conclusions}.

Throughout this paper we adopt a standard $\Lambda$CDM cosmology with
$\left[ H_0, \Omega_{\mathrm{m}}, \Omega_{\Lambda}, \sigma_8
\right]=\left[70\,\mathrm{km\,s^{-1}\,Mpc^{-1}},0.3,0.7,0.9\right]$.
We use Vega magnitudes if not stated otherwise.

\section{The data and the samples}
\label{sec:data}
\subsection{DPS images}
The ESO Deep Public Survey is a deep multicolour survey carried out
with the Wide Field Imager (WFI), an eight-chip CCD camera of
$34\arcmin\times33\arcmin$ field-of-view mounted at the MPG/ESO2.2\,m
telescope at La Silla. The images used for this study are described in
\cite{2006A&A..Hildebrandt} along with the characteristics of the WFI
filter-set. Furthermore, details on the raw data, the data reduction
with our THELI pipeline \citep{2005AN....326..432E}, the astrometric
and photometric calibration, the quality control, and the data release
to the scientific community can be found there.

For our studies on LBGs we use seven fields (2\,sq.\,deg) with
complete coverage in the $UBVRI$-filters, in particular the fields
Deep1a, 1b, 2b, 2c, 3a, 3b, and 3c, respectively. Their properties are
summarised in Table~\ref{tab:fields}.

\begin{table*}
  \caption{The DPS fields with five colour coverage. The limiting magnitudes in columns four to eight are measured in a circular aperture of $2\times$FWHM diameter from the $1\sigma$ sky background fluctuations of the images before convolution with a Gaussian filter (see text). In column nine the seeing FWHM values after convolution are given. Column ten contains the number of objects satisfying Eq.~(\ref{equ:U_dropout_selection}). In the fields Deep2b and Deep3a less $U$-dropouts are selected due to inferior quality of the imaging data (see text). The completeness limits for LBG selection in column eleven are estimated from Fig.~\ref{fig:numbercounts}. These visual estimates are very rough with an accuracy of $\sim0.5\,\mathrm{mag}$. The effective area used for LBG selection is given in the last column.}
\label{tab:fields}
\begin{tabular}{l r r r r r r r c r r r}
  \hline
  \hline
  field & RA [h m s] & Dec [d m s] & \multicolumn{5}{c}{1-$\sigma$ mag lim. [Vega mags]} & conv. & $N_{\mathrm{LBG}}$ & LBG compl. limit & eff. area\\
  & J2000.0    & J2000.0     & $U$ & $B$ & $V$ & $R$ & $I$ & seeing & & $R$ & [$\mathrm{arcmin}^2$]\\
  \hline
  Deep1a& 22:55:00.0 & $-$40:13:00 & 27.0 & 27.8 & 27.5 & 27.4 & 26.3 & $1\farcs3$ & 1420 & 25.0 & 1045\\
  Deep1b& 22:52:07.1 & $-$40:13:00 & 27.0 & 27.5 & 27.1 & 27.2 & 26.2 & $1\farcs3$ & 1114 & 24.5 & 1036\\
  Deep2b& 03:34:58.2 & $-$27:48:46 & 26.8 & 28.0 & 27.5 & 26.9 & 26.3 & $1\farcs3$ & 492  & 24.0 & 1025\\
  Deep2c& 03:32:29.0 & $-$27:48:46 & 27.1 & 29.0 & 28.6 & 28.5 & 26.3 & $1\farcs0$ & 2181 & 25.5 & 1064\\
  Deep3a& 11:24:50.0 & $-$21:42:00 & 26.6 & 27.9 & 27.2 & 27.3 & 25.8 & $1\farcs1$ & 456 & 24.0 & 1033\\
  Deep3b& 11:22:27.9 & $-$21:42:00 & 26.9 & 27.9 & 27.3 & 27.3 & 26.2 & $1\farcs0$ & 1484 & 25.0 & 1072\\
  Deep3c& 11:20:05.9 & $-$21:42:00 & 27.0 & 28.1 & 27.3 & 27.2 & 25.8 & $1\farcs0$ & 1679 & 25.0 & 998\\
  \hline
  &&&&&&&&& $\Sigma=8826$ & & $\Sigma=7273$\\
\end{tabular}
\end{table*}

Note that the images of the field Deep2c used in the current study are
different from the images used in \cite{2005A&A....Hildebrandt}. More
data have become available so that the current images in this field
are considerably deeper.

\subsection{Catalogue extraction}
First, the different colour images of one field are trimmed to the
same size. The seeing is measured for these images and SExtractor
\citep{1996A&AS..117..393B} is used to create RMS maps.
From these RMS images the local limiting magnitude ($1\sigma$ sky
background in a circular aperture of $2\times$FWHM diameter) in each
pixel is calculated and limiting magnitude maps are created. Then
every image is convolved with an appropriate Gaussian filter to match
the seeing of the image with the worst seeing value.

For the catalogue extraction SExtractor is run in dual-image mode with
the unconvolved $R$-band image for source detection and the convolved
images in the five bands for photometric flux measurements. In this
way it is assured that the same part of a galaxy is measured in every
band. In the absence of a strong spatial colour gradient in an object
this method should lead to unbiased colours especially for
high-redshift objects of small apparent size. In the following, we use
isophotal magnitudes when estimating colours or photometric redshifts
of objects. The total $R$-band magnitudes always refer to the
SExtractor parameter MAG\_AUTO measured on the unconvolved $R$-band
image.

\subsection{Photometric redshift estimation}
Photometric redshifts are estimated for all objects in the catalogue
from their $UBVRI$ photometry using the publicly available code
\emph{Hyperz}~\citep{2000A&A...363..476B}. The technique applied is
essentially the same as described in \cite{2005A&A....Hildebrandt}
with the only difference that in the current study we use isophotal
magnitudes extracted from images with matched seeing instead of seeing
adapted aperture magnitudes. In Hildebrandt et al. (2006, in
preparation) we compare our photometric redshift estimates to several
hundred spectroscopic redshifts from the VIMOS VLT Deep Survey
\citep[VVDS; ][]{2004A&A...428.1043L} in the field Deep2c and find
that the combination of isophotal magnitudes with the templates
created from the library of \cite{1993ApJ...405..538B} yields the
smallest scatter and outlier rates at least for the redshifts probed
by the VVDS ($z<1.4$). For $R<24$ we find a standard deviation of
$\sigma=0.052$ for the quantity $\Delta
z=\left(z_{\mathrm{spec}}-z_{\mathrm{phot}}\right)/\left(1+z_{\mathrm{spec}}\right)$
after rejecting 5\% of outliers (objects with $\Delta z>0.15$).  With
increasing redshift and decreasing angular size of the objects the
isophotal magnitudes on images with matched seeing should approach the
seeing adapted aperture magnitudes.  Investigating the redshift
distributions of our $U$-dropout sample (see below) we see no
significant difference between the two approaches so that the results
of the clustering measurements are not influenced by this choice.

\subsection{Sample selection}
\label{sec:selection}
\begin{figure}
\resizebox{\hsize}{!}{\includegraphics{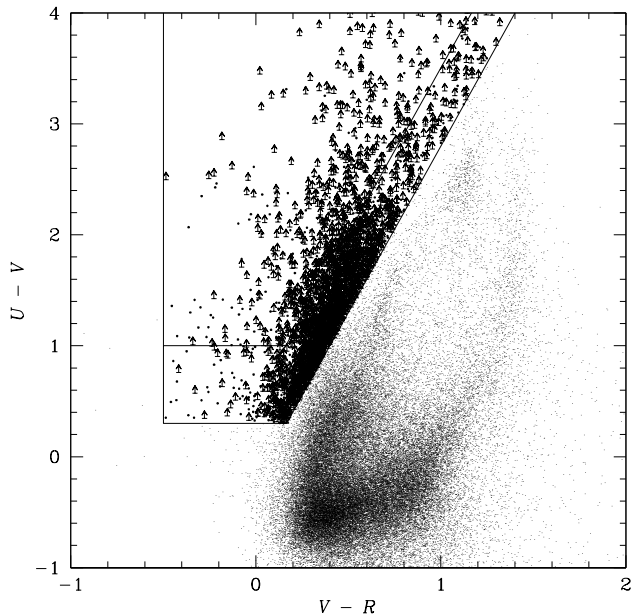}}
\caption{\label{fig:selection}$(U-V)$ vs. $(V-R)$ colour-colour
  diagram of galaxies in the field Deep2c. The upper box represents
  the old selection criteria adopted in \cite{2005A&A....Hildebrandt}
  while the lower extension represents the selection criteria of the
  current study. The $(U-V)$ colours of objects that are satisfying
  the LBG selection criterea but are not detected in the $U$-band are
  lower limits (\emph{arrows}). In contrast the detected objects
  satisfying the selection criterea are plotted as \emph{filled
    circles}.}
\end{figure}

In \cite{2005A&A....Hildebrandt} we chose quite conservative criteria
for our $U$-dropout selection. In particular, we tried to define
colour cuts in such a way to avoid regions in colour space with a
considerable amount of contamination. Supported by our simulations
(see Sect.~\ref{sec:simulations}), and after refining our photometric
measurements (see above), which should yield better colours,
especially for the possible contaminants, we decided to relax our
selection criteria. The performance of our selection in terms of
contamination and efficiency is analysed in
Sect.~\ref{sec:simulations} by means of simulated colour catalogues.
In Fig.~\ref{fig:selection} the colour-colour diagram for one field is
shown with the old and the new selection criteria represented by the
boxes. We select candidates for $z=3$ LBGs in the following way:
\begin{eqnarray}
  \label{equ:U_dropout_selection}
  0.3 &<&(U-V) \; , \nonumber \\
  -0.5 &<& (V-R)<1.5 \; ,\\
  3(V-R) &<& (U-V)+0.2 \; , \nonumber
\end{eqnarray}

Furthermore, we require every candidate to be detected in $V$ and $R$
and to be located in a region on our images where the local limiting
magnitudes in $UVR$ are not considerably decreased. This is necessary
to avoid complex selection effects in colour space due to varying
depths over a single field. In our catalogues we find 8826 objects
satisfying these selection criteria.

Every candidate is inspected visually in all five filters. Moreover,
the redshift-probability distribution
\citep[see][]{2005A&A....Hildebrandt}, the location in the field, and
the location in the colour-colour diagram is checked for every object.
Approximately one third of the candidates is rejected in this way.
Most of these rejected objects are influenced by the straylight from
bright neighbouring objects (indicated by their extraction flags or
visible in the $10''\times10''$ thumbnail images) so that their
colours cannot be trusted. Merely 72 objects ($<1\%$) are rejected due
to their redshift-probability distribution in combination with a
suspicious location in the colour-colour diagram near the stellar
locus. Thus, the photometric redshift distribution is almost not
affected by the exclusion of these objects.

\begin{figure*}
\resizebox{\hsize}{!}{\includegraphics{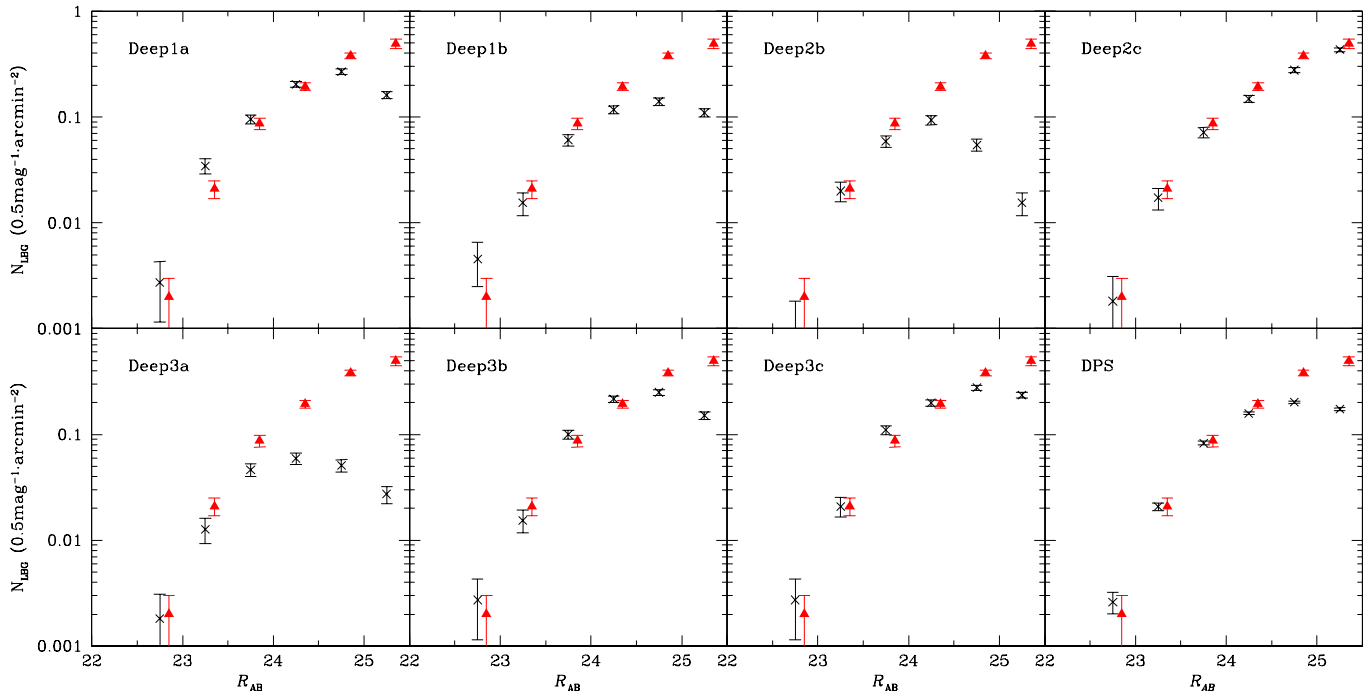}}
\caption{\label{fig:numbercounts}Source counts of LBGs as a function
  of $R$-band magnitude. The DPS densities are represented by crosses
  while the \cite{1999ApJ...519....1S} densities are represented by
  triangles which are offset by +0.1\,mag just for clarity.}
\end{figure*}

The magnitude dependent angular number-densities of the accepted
candidates for the seven fields and for the whole DPS are displayed in
Fig.~\ref{fig:numbercounts} in comparison to the values found by
\cite{1999ApJ...519....1S}. For $R<24$ all seven fields show
approximately the same LBG source density within a reasonable
field-to-field variance. The field Deep2b can be regarded as fairly
complete down to $R=24.5$ and the fields Deep1a, 3b, and 3c,
respectively, down to $R=25$. But only the field Deep2c shows the same
density as the survey by \cite{1999ApJ...519....1S} down to $R=25.5$.
The underdensity of Deep1b and Deep2b can be explained by the inferior
seeing in the detection images ($R$) while the underdensity of Deep3a
is due to a shallower $U$-band image.

As long as one restricts investigations in a given magnitude bin to
fields that can be regarded as uniform in terms of LBG selection,
these investigations are not subject to systematic effects due to
varying selection efficiency between the fields. This is in particular
important in the clustering analysis to avoid artificial correlations
originating from non-uniform depths.

\begin{figure}
  \resizebox{\hsize}{!}{\includegraphics{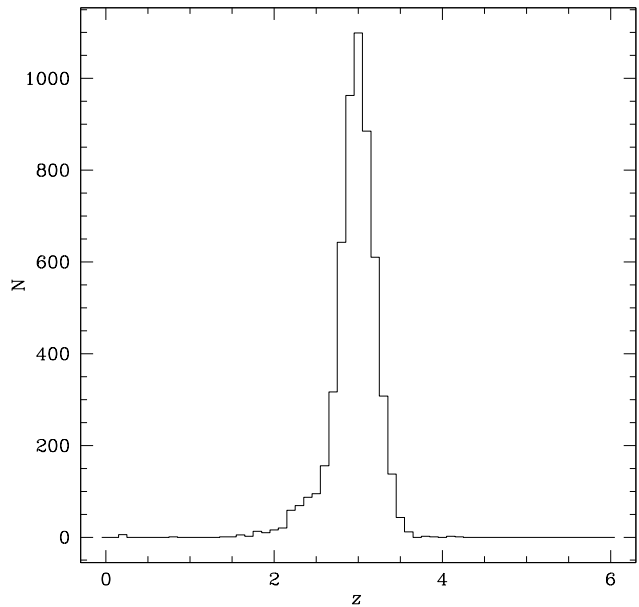}}
\caption{\label{fig:phot_z_dist}Photometric redshift distribution of
  all accepted LBG candidates. The distribution has a mean of
  $\left<z\right>=2.96$ and an RMS of $\sigma=0.24$.}
\end{figure}

The photometric redshift distribution of all accepted objects is shown
in Fig.~\ref{fig:phot_z_dist}. The mean redshift of
$\left<z\right>=2.96$ (4-$\sigma$ outliers are rejected) is
essentially the same as the spectroscopic mean found by
\cite{2005ApJ...619..697A} for their LBG sample, but our distribution
seems to be slightly narrower ($\sigma=0.24$).  However, without
spectroscopic information we are not able to judge whether this is due
to the different selection method with a different camera and filter
set or due to imperfect photometric redshift estimation. In the
following we will use the distribution inferred from our photometric
redshifts indicating whenever the width has a large influence on our
results.

\section{Simulations of objects' colours in the DPS}
\label{sec:simulations}
Without a large spectroscopic survey of our LBGs at hand the only
possible way to estimate properties of our LBG samples like, e.g., the
contamination rate is to create a simulated colour catalogue.

\subsection{The role of stars in our sample}
\begin{figure}
\resizebox{\hsize}{!}{\includegraphics{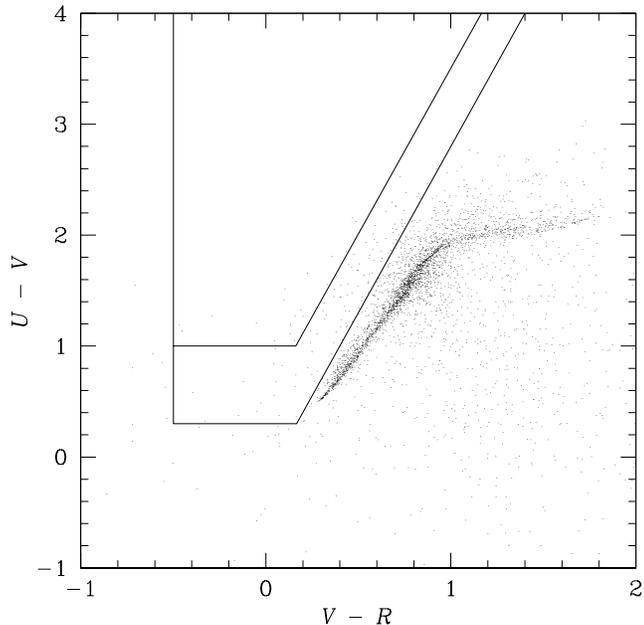}}
\caption{\label{fig:UVR_stars}$(U-V)$ vs. $(V-R)$ colour-colour
  diagram of simulated stars in the field Deep1a. The boxes are the
  same as in Fig.~\ref{fig:selection}.}
\end{figure}

The TRILEGAL galactic model by \cite{2005A&A...436..895G} is used to
simulate the number of stars in all seven fields and their colours in
the WFI filter set. In this way we obtain accurate $U-V$ and $V-R$
colours and are able to quantify the amount of stellar contamination
in the LBG selection box. In Fig.~\ref{fig:UVR_stars} the colours of
stars in the field Deep1a are shown, representative for the whole
survey. The selection criteria were chosen in such a way that stellar
contamination is very low, which is confirmed by
Fig.~\ref{fig:UVR_stars} (see also the bottom panel of
Fig.~\ref{fig:segregation}).

\subsection{Colours of galaxies}
\label{sec:sim_gal}
\begin{figure}
\resizebox{\hsize}{!}{\includegraphics{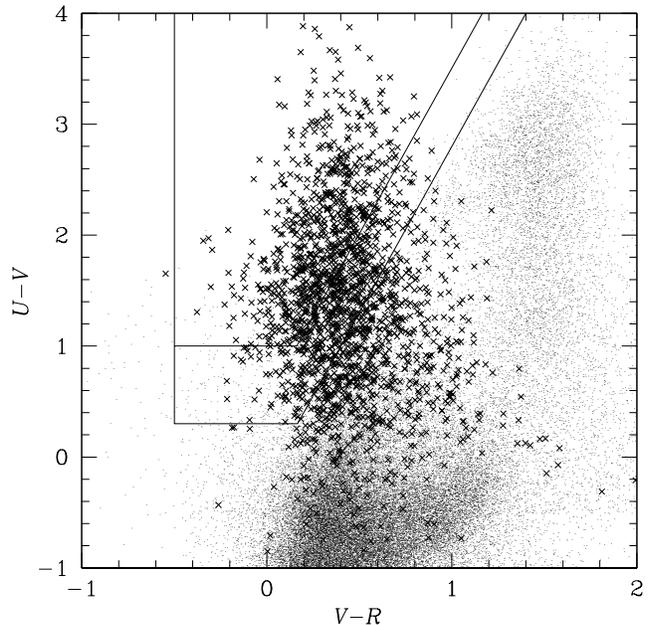}}
\caption{\label{fig:UVR_gal_sim}$(U-V)$ vs. $(V-R)$ colour-colour
  diagram of simulated galaxies. The boxes are the same as in
  Fig.~\ref{fig:selection}. Objects with a redshift of $z>2.5$ are
  plotted as crosses. We find good agreement in the overall shape of
  the colour distribution of the simulated galaxies to the real data
  (see Fig.~\ref{fig:selection}). The slight differences, however, may
  be attributed to an imperfect template set and to a redshift
  distribution which is subject to cosmic variance.}
\end{figure}

The code \emph{Hyperz} cannot only be used to estimate photometric
redshifts but also to create colour catalogues of galaxies at
different redshifts and of different spectral types. We simulate huge
random mock catalogues in magnitude bins of width 0.5\,mag of 500\,000
galaxies each, evenly distributed over the redshift interval $0<z<7$
and over all spectral types from the library of
\cite{1993ApJ...405..538B} provided by \emph{Hyperz}.  Magnitude
errors are simulated by \emph{Hyperz} according to the
\mbox{1-$\sigma$} limits in the five bands. This is done separately
for every field taking into account the different depths in the five
bands. From the photometric redshift code \emph{BPZ}
\citep{2000ApJ...536..571B} we extract magnitude- and spectral
type-dependent redshift distributions derived from the Hubble Deep
Field. We assign the two reddest \emph{Hyperz} spectral types from the
library of \cite{1993ApJ...405..538B} (``Burst'' and ``E'') to the
\emph{BPZ} $z$-distribution for elliptical galaxies, two intermediate
types (``Sb'' and ``Sc'') to the $z$-distribution for spirals, and the
two bluest spectral types (``Sd'' and ``Im'') to the $z$-distribution
for star-forming galaxies. Galaxies are taken from the evenly
distributed catalogues with numbers and spectral types according to
these redshift distributions to create realistic catalogues for
0.5\,mag intervals. Finally, from these catalogues the galaxies are
taken with numbers scaled to the $I$-band number-counts in the seven
fields.  In this way, for every field a catalogue is created which
gives a fair representation of our data in 0.5\,mag wide intervals.
The $U-V$ vs.  $V-R$ colour-colour diagram of the simulated galaxies
is shown in Fig.~\ref{fig:UVR_gal_sim}.

Certainly, it is a strong assumption that the chosen template set
represents the galaxy population in our data at all redshifts.
Furthermore, the redshift distributions were extracted from the Hubble
Deep Field which is subject to cosmic variance. These shortcomings are
most probably responsible for the slight differences between
Fig.~\ref{fig:selection} and Fig.~\ref{fig:UVR_gal_sim} and the
simulations should only be regarded as rough estimates.

Applying the colour cuts from Eq.~(\ref{equ:U_dropout_selection}) to
the simulated star- and galaxy-catalogues we obtain estimates for the
contamination rate in our LBG sample at different magnitudes which are
plotted in the bottom panel of Fig.~\ref{fig:segregation} and included
in Table~\ref{tab:clustering}. For the considered magnitude range of
$22.5<R<26$ the total contamination is below 20\%.

\begin{figure}
\resizebox{\hsize}{!}{\includegraphics{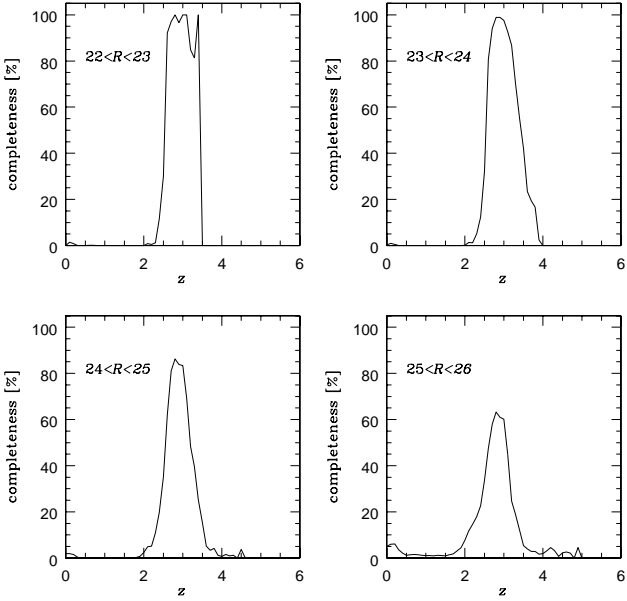}}
\caption{\label{fig:completeness} Completeness as defined in the text
  of our LBG selection in dependence of magnitude and redshift
  averaged over the seven fields.}
\end{figure}

We define the completeness in a particular magnitude- and redshift-bin
as the ratio of the number of objects selected by our criteria to the
number of objects in the whole catalogue. Since the objects in our
mock catalogue are generated in such a way that their magnitude errors
are derived from the typical limiting magnitudes in the DPS, with
these simulations the completeness of our selection can only be
quantified with respect to this catalogue. In
Fig.~\ref{fig:completeness} the completeness in dependence of $R$-band
magnitude and redshift is shown.  The values should be regarded as an
upper bound for the total completeness with respect to the whole
galaxy population since some of them may be entirely undetectable in
our images due to low surface brightness.  Certainly, this becomes
more serious for $R>25$ where our LBG number-counts start to drop (see
Fig.~\ref{fig:numbercounts}).

\begin{figure}
  \resizebox{\hsize}{!}{\includegraphics{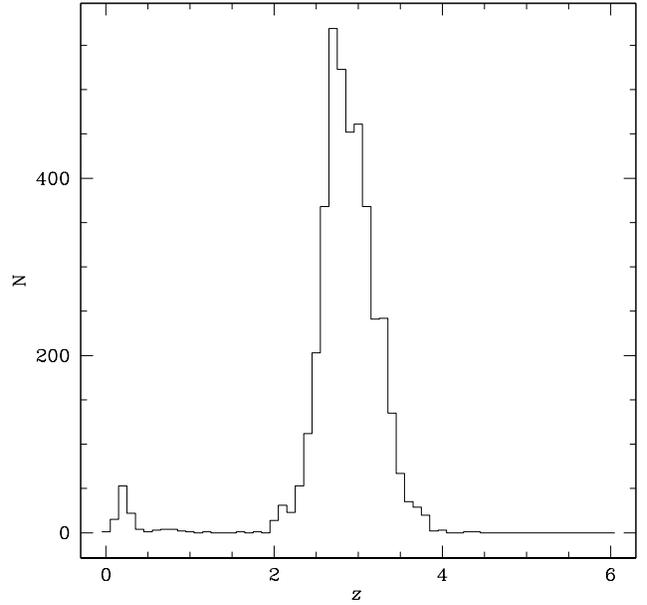}}
  \caption{\label{fig:z_dist_sim}Redshift distribution of $23<R<25$
    (to avoid magnitude regions with higher contamination) galaxies
    selected from the mock catalogues of Deep1a, 2c, 3b, and 3c by
    Eq.~(\ref{equ:U_dropout_selection}) showing good agreement to
    Fig.~\ref{fig:phot_z_dist} with only a slightly wider peak. The
    distribution of galaxies with $2<z<4$ has a mean of
    $\left<z\right>=2.89$ and a standard deviation of $\sigma=0.30$.}
\end{figure}

The redshift distribution of the selected simulated galaxies in the
magnitude interval $23<R<25$ is shown in Fig.~\ref{fig:z_dist_sim}
with good agreement to the photometric redshift distribution in
Fig.~\ref{fig:phot_z_dist}.

\section{Clustering properties}
\label{sec:clustering}
\subsection{Method}
\label{sec:method}
The angular correlation function is estimated as described in
\cite{1993ApJ...412...64L},

\begin{equation}
\label{eq:Landy}
\omega(\theta)=\frac{\mathrm{DD-2DR+RR}}{\mathrm{RR}}\,.
\end{equation}
The numbers of galaxy pairs with a separation between $\theta$ and
$\theta+\delta\theta$ in the data (DD), in a random catalogue with the
same field geometry and density (RR), and between the data and the
random catalogue (DR) are counted on the DPS fields with comparable
selection efficiency for a particular magnitude range separately.  The
angular correlation function for the whole survey is then estimated
from the sums of the three quantities over the fields.  We apply
Poissonian errors for the angular correlation function
\citep{1993ApJ...412...64L},

\begin{equation}
\delta\omega(\theta)=\sqrt{\frac{1+\omega(\theta)}{\mathrm{DD}}}\,.
\end{equation}
Although the area of the DPS is rather large in comparison to previous
$U$-dropout surveys the results may nevertheless be subject to cosmic
variance. We estimate the amplitude of the cosmic variance from the
field-to-field variance which includes cosmic variance as well as
shot-noise from the limited number of galaxies used for the estimate
of $\omega(\theta)$. We find values which are comparable in size to
the Poissonian errors. This means that the errors are dominated by
shot-noise while cosmic variance is negligible. Thus, the application
of Poissonian errors is justified.

\begin{figure}
\resizebox{\hsize}{!}{\includegraphics{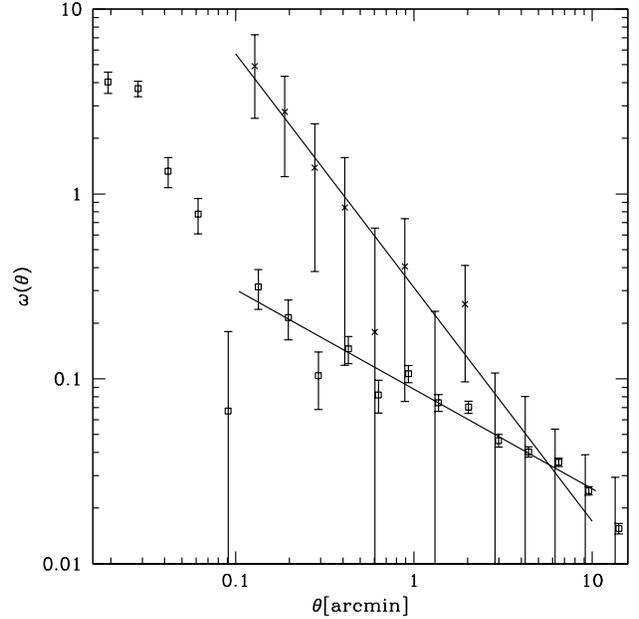}}
\caption{\label{fig:acf_tot}Angular correlation function for
  $U$-dropouts with $22.5<R<23.5$ (crosses) of all seven fields and with
  $22.5<R<26$ (open squares, slightly offset for clarity) of the
  fields Deep1a, Deep2c, Deep3b, and Deep3c. The solid lines represent
  power law fits to the data in the range
  $0.1\arcmin<\theta<10\arcmin$. The angular correlation function of
  the faint sample shows an excess on small scales with respect to the
  power law fitted to the data on intermediate scales.}
\end{figure}

In Fig.~\ref{fig:acf_tot} the angular correlation function is shown
for all $U$-dropouts with $22.5<R<23.5$ of all seven fields as well as
with $22.5<R<26$ of the fields Deep1a, Deep2c, Deep3b, and Deep3c.

A power law, 
\begin{equation}
\label{eq:power_law}
\omega(\theta)=A_{\omega}\theta^{-\delta}\,, 
\end{equation}
is fitted to the angular correlation function and the Limber equation
\citep[see][]{2005A&A....Hildebrandt} is used to estimate the
real-space correlation function, $\xi$.\footnote{Note that the Limber
  equation is inaccurate to some degree due to the relatively narrow
  distribution in comoving distance of our LBGs. See
  \cite{2006astro-ph...0609165...v1} for details. This is certainly
  also the case for other LBG studies applying the Limber equation so
  that the relative comparisons presented here are not affected
  seriously.} In this step we apply our photometric redshift
distribution presented in Sect.~\ref{sec:selection}. We parametrise
the power law approximation of the real-space correlation function in
the following form:

\begin{equation}
\xi\left(r\right)=\left( \frac{r}{r_0} \right)^{-\gamma}\,,
\end{equation}
with $r$ being the comoving distance, $r_0$ being the comoving
correlation length, and $\gamma=1+\delta$.

The estimator from Eq.~(\ref{eq:Landy}) is known to be biased low
because the galaxy density in the field is estimated from the data
itself and no fluctuations on the scale of the field size are
accounted for,

\begin{equation}
\omega_{\mathrm{real}}(\theta)=\omega(\theta)+\mathrm{IC}\,,
\end{equation}
with the bias $\mathrm{IC}$ usually called ``the integral
constraint''.

It can be shown \citep[see e.g.][]{2005ApJ...619..697A} that the
expectation value of this bias equals the variance of galaxy-density
fluctuations on the size of the field-of-view. We estimate the
integral constraint by the method outlined in
\cite{2005ApJ...619..697A} from the linear cold dark matter (CDM)
power spectrum. The variance of mass $\sigma_\mathrm{CDM}^2$ in our
typical survey volumes can be estimated from integrating the power
spectrum over the Fourier transform of such a survey volume. The
survey volume of $z=3$ LBGs in a single DPS field can be reasonably
approximated by a square on the sky (comoving dimensions of
$42\times42 (h^{-1}\mathrm{Mpc})^2$ at redshift $z=3$) and a Gaussian in
radial direction ($\sigma=88\,h^{-1}\mathrm{Mpc}$) resulting in
$\sigma_\mathrm{CDM}^2=0.0017$.

Assuming a linear relationship between the fluctuations of the mass
density and the galaxy density, the linear bias factor can be
estimated from the correlation function. An iterative approach to
estimate the IC first and then the bias factor from the fitted
real-space correlation function converges quickly. For a detailed
description of the method we refer to \cite{2005ApJ...619..697A}.

\subsection{Results}
\begin{figure}
  \resizebox{\hsize}{!}{\includegraphics[angle=-90]{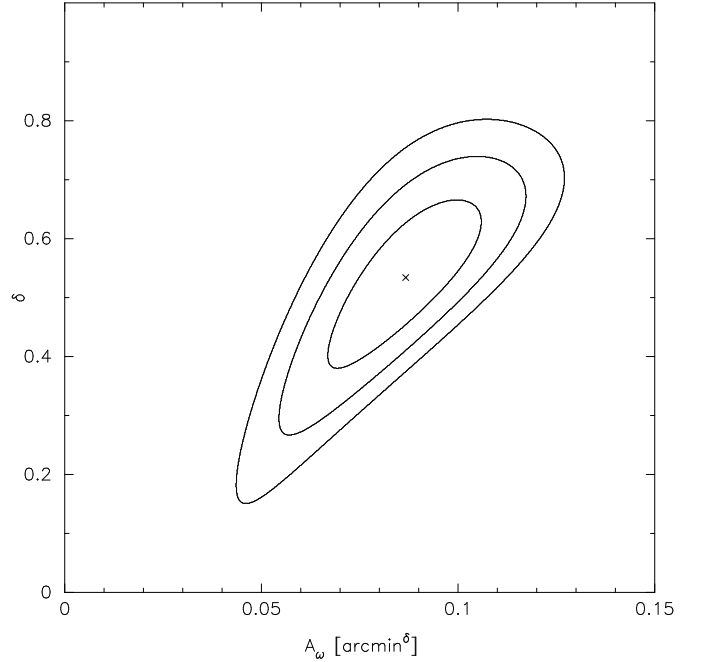}}
  \caption{\label{fig:fit_contours}Joint 68.3\%-, 95.4\%-, and
    99.8\%-confidence regions (corresponding to $\Delta\chi^2=\left[2.3;
    6.2; 11.8\right]$) for the power law parameters $A_\omega$ and $\delta$
    (see Eq.~(\ref{eq:power_law})) in the fit to the angular correlation
    function of the $22.5<R<26$ subsample.}
\end{figure}

In Table~\ref{tab:clustering} the results for various subsamples of
the LBGs are presented in comparison to results from previous studies.
The errors on the correlation lengths are derived from Monte-Carlo
simulations taking into account the fitting errors for $A_{\omega}$
and $\delta$ and assuming that these are Gaussian and uncorrelated. In
Fig.~\ref{fig:fit_contours} the confidence regions for $A_{\omega}$
and $\delta$ for the $22.5<R<26$ subsample are plotted as
2-dimensional contours.

We investigate the influence of the binning by estimating the
correlation length for 10 to 25 bins and find that the standard
deviation over these 16 binnings is comparable or smaller than the
error introduced by the fitting. We choose a common binning for all
magnitude intervals which samples the correlation function well.

There are, however, systematic uncertainties in our correlation
analysis, the most serious being our selection function. We must rely
on the validity of our photometric redshift distribution shown in
Fig.~\ref{fig:phot_z_dist}.  The derived correlation lengths certainly
depend on the width of this distribution, with wider distributions
resulting in larger correlation lengths. Moreover, the results from
the simulations in Sect.~\ref{sec:simulations} may be subject to
cosmic variance since the underlying redshift distributions were
derived from the small HDF. Thus, our contamination and completeness
estimates must be regarded as approximate values.

\begin{table*}
  \caption{Clustering measurements of LBGs in the DPS fields for different limiting magnitudes and results from other surveys for comparison. O2005 refers to \cite{2005ApJ...635L.117O} and A2005 refers to \cite{2005ApJ...619..697A}. The power law fits to the angular correlation function are performed in the range $0\farcm1\le \theta \le 10\arcmin$. The errors on the correlation lengths are estimated from the fitting errors of the slope and the amplitude of the angular correlation function. No possible systematical uncertainties introduced by the photometric redshift distribution, the binning, etc. are included. Notice that the $22.5<R<25.5$, $22.5<R<26$, and $23.3<R<25.3$ samples suffer from some incompleteness at the faint end which may be the reason for non-evolution from $\mathrm{mag_{lim}}=25$ to $\mathrm{mag_{lim}}=26$ and slight disagreement to the results by \cite{2005ApJ...619..697A}. In the second column the number of fields are listed with ``7'' corresponding to all seven DPS fields with $UBVRI$ coverage, ``5'' corresponding to the fields Deep1a, 1b, 2c, 3b, and 3c, and ``4'' corresponding to Deep1a, 2c, 3b, and 3c, respectively. The values for the integral constraint, IC, and the linear bias factor, $b$, in the sixth and seventh column are estimated as described in Sect.~\ref{sec:method} and the contamination fractions, $f$, in column eight are derived from the simulations presented in Sect.~\ref{sec:simulations}. The average number of LBGs per halo, $\left<N_{\mathrm{g}}\right>$ , and the average mass of an LBG hosting halo, $\left<M_{\mathrm{halo}}\right>$ are estimated as detailed in Sect.~\ref{sec:halo}. Note that the $i'_{\mathrm{AB},z=4}$ limiting magnitudes in the study by \cite{2005ApJ...635L.117O} can be related to our $R_{\mathrm{Vega}}$ limiting magnitudes by $i'_{\mathrm{AB},z=4,\mathrm{lim}}\hat=R_{\mathrm{Vega,lim}}+1$ as described in the text.}
\label{tab:clustering}
\begin{tabular}{c c r r r r r r r r}
  \hline
  \hline
  Sample & No. fields & N & \multicolumn{1}{c}{$\gamma$} & $r_0$ & \multicolumn{1}{c}{$\mathrm{IC}$} & $b$ & $f$ & $\left<N_{\mathrm{g}}\right>$ & $\log\left<M_{\mathrm{halo}}\right>$ \\
  &&&& [$h^{-1}\mathrm{Mpc}$] &&&[\%] &&[$h^{-1}M_{\sun}$]\\
  \hline
  $22.5\le R \le 23.5$ & 7 & 228  & $2.26\pm0.20$ & $7.2\pm1.2$ & 0.045 & 5.1 &18.4 & $0.46\pm0.73$ & $12.79^{+0.05}_{-0.05}$\\
  $22.5\le R \le 24.0$ & 7 & 965  & $1.92\pm0.09$ & $6.3\pm0.6$ & 0.015 & 3.0 &11.1 & $0.43\pm0.37$ & $12.46^{+0.03}_{-0.04}$\\
  $22.5\le R \le 24.5$ & 5 & 1864 & $1.55\pm0.08$ & $5.2\pm0.4$ & 0.015 & 3.0 & 9.3 & $0.42\pm0.24$ & $12.29^{+0.09}_{-0.10}$\\
  $22.5\le R \le 25.0$ & 4 & 2950 & $1.57\pm0.06$ & $4.8\pm0.3$ & 0.013 & 2.8 &12.0 & $0.61\pm0.51$ & $12.12^{+0.10}_{-0.12}$\\
  $22.5\le R \le 25.5$ & 4 & 3913 & $1.58\pm0.05$ & $4.8\pm0.3$ & 0.014 & 2.9 &15.6 & $0.44\pm0.23$ & $12.15^{+0.15}_{-0.24}$\\
  $22.5\le R \le 26.0$ & 4 & 4363 & $1.54\pm0.04$ & $4.8\pm0.2$ & 0.014 & 2.9 &19.4 & $0.50\pm0.24$ & $12.15^{+0.14}_{-0.20}$\\
  \hline
  O2005 $(i'_{\mathrm{AB},z=4}<24.5)$&$-$& 239  & $2.1\pm0.4$ & $4.9^{+4.3}_{-4.1}$ & $-$ & $-$ & $-$       & $0.2^{+0.2}_{-0.2}$ & $12.3^{+0.1}_{-0.6}$\\
  O2005 $(i'_{\mathrm{AB},z=4}<25.0)$&$-$& 808  & $1.9\pm0.3$ & $5.5^{+1.7}_{-2.1}$ & $-$ & $-$ & $-$       & $0.3^{+0.4}_{-0.3}$ & $12.3^{+0.1}_{-0.2}$\\
  O2005 $(i'_{\mathrm{AB},z=4}<25.5)$&$-$& 2231  & $1.8\pm0.1$ & $5.0^{+0.7}_{-0.8}$ & $-$ & $-$ & $-$      & $0.6^{+0.1}_{-0.5}$ & $12.1^{+0.1}_{-0.1}$\\
  O2005 $(i'_{\mathrm{AB},z=4}<26.0)$&$-$& 4891 & $1.8\pm0.1$ & $5.0^{+0.4}_{-0.4}$ & $-$ & $-$ & $-$       & $0.6^{+0.1}_{-0.1}$ & $12.0^{+0.1}_{-0.1}$\\
  O2005 $(i'_{\mathrm{AB},z=4}<26.5)$&$-$& 8639 & $1.6\pm0.1$ & $4.8^{+0.2}_{-0.3}$ & $-$ & $-$ & $-$       & $0.6^{+0.1}_{-0.1}$ & $11.9^{+0.05}_{-0.05}$\\
  O2005 $(i'_{\mathrm{AB},z=4}<27.0)$&$-$& $12\,921$ & $1.6\pm0.1$ & $4.4^{+0.1}_{-0.2}$ & $-$ & $-$ & $-$  & $0.6^{+0.1}_{-0.2}$ & $11.8^{+0.07}_{-0.04}$\\
  \hline
  $23.3\le R \le 25.3$ &4& 3541 & $1.60\pm0.03$ & $5.0\pm0.2$ & 0.015 & 3.0 &  18.7 & $-$ & $-$\\
  \hline
  A2005 $(23.5<R_{\mathrm{AB}}<25.5)$&$-$&$-$& $1.57\pm0.14$ & $4.0\pm0.6$ & $\sim0.01$ & $-$ & $-$ & $-$ & $-$\\
  \hline
\end{tabular}
\end{table*}

Furthermore, we cannot correct our clustering measurements for
contamination directly as there are no spectroscopic observations of
our WFI-selected LBG samples available yet. In general, a
contamination rate $f$ of uncorrelated sources in our catalogues will
lead to an angular correlation function with a measured amplitude
$A=(1-f)^2\,A_\mathrm{real}$ implying a corrected correlation length
$r_{0,\mathrm{corr}}=(1-f)^{-2/\gamma}r_0$. However, as we do not know
from our simulations about the exact clustering behaviour of the
contaminants we do not apply such a correction.

\begin{figure}
  \resizebox{\hsize}{!}{\includegraphics{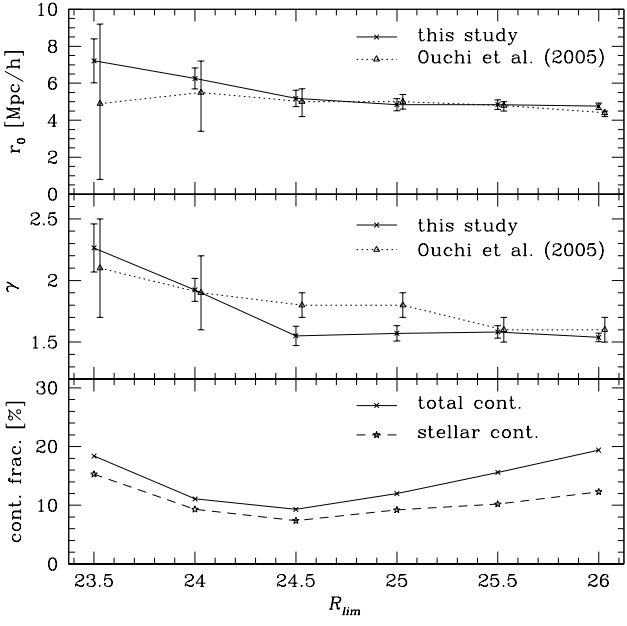}}
  \caption{\label{fig:segregation}Dependence of the correlation length
    and the slope of the correlation function on $R$-band limiting
    magnitude (\emph{upper} and \emph{middle} panels, respectively).
    The crosses and the solid lines represent our data. The triangles
    (slightly offset for clarity) and the dotted lines show the values
    from \cite{2005ApJ...635L.117O} at $z=4$ in comparison. We relate
    their $i'_{\mathrm{AB}}$ limiting magnitudes to our
    $R_\mathrm{Vega}$ limiting magnitudes as described in the text.
    The lower panel shows the contamination fraction for the different
    magnitude-limited subsamples with $22.5<R<R_{\mathrm{lim}}$ with
    the solid line representing the total contamination and the dashed
    line representing the stellar contamination.}
\end{figure}

We see clustering segregation with rest-frame UV luminosity in our
data. In Fig.~\ref{fig:segregation} the dependence of the correlation
lengths and the slope of the correlation function are plotted against
limiting magnitude along with the contamination estimates from
Sect.~\ref{sec:sim_gal}. The correlation lengths for the different
subsamples decrease monotonically with limiting magnitude down to
$R_{\mathrm{lim}}=25$ and then stay constant whereas the slope
decreases down to $R_{\mathrm{lim}}=24.5$.

The observation that more luminous LBGs show larger correlation
lengths was reported quite some time ago \citep[see e.g.
][]{2001ApJ...550..177G, 2001ApJ...558L..83O}. In the CDM framework
more massive halos are more strongly biased than less massive halos
with respect to the whole mass distribution. Thus, brighter LBGs are
supposed to be hosted by more massive halos than fainter ones. For a
quantitative analysis of halo properties see Sect.~\ref{sec:halo}.

We compare our results to precise recent measurements of $z=4$ LBG
clustering by \cite{2005ApJ...635L.117O}. For these comparisons our
$R_{\mathrm{Vega}}$ limits must be converted to the AB system
($+0.2\,\mathrm{mag}$) and the distance modulus between $z=3$ and
$z=4$ must be added ($+0.8\,\mathrm{mag}$ for $\Lambda$CDM).  Since
the $R$-band at $z=3$ closely resembles the $I$-band at $z=4$ in terms
of restframe wavelength coverage we do not apply a $k$-correction. The
agreement of both studies shown in Table~\ref{tab:clustering} and
Fig.~\ref{fig:segregation} is excellent with most corresponding
measurements lying within the \mbox{1-$\sigma$} intervals. Considering
the systematic differences introduced by different filter-sets,
different depths, different selection criteria, etc., the agreement is
rather impressive. However, considering the cosmic time and the
structure formation that took place between $z=4$ and $z=3$ this means
that an LBG at $z=3$ is hosted by a significantly more massive halo
than an LBG of the same luminosity at $z=4$ (see
Sect.~\ref{sec:halo}).

\cite{2005ApJ...619..697A} confine their samples of optically
selected, star-forming galaxies to the magnitude range $23.5\le
R_\mathrm{AB}\le25.5$, which corresponds to $23.3\le R_\mathrm{Vega}
\le 25.3$. In this magnitude range the correlation length derived from
our survey is slightly larger than the value found by
\cite{2005ApJ...619..697A} while the slopes found in both studies
agree very well within the uncertainties (see
Table~\ref{tab:clustering}). To compare the depth of our images with
the ones used in \cite{2005ApJ...619..697A} we calculate
\mbox{1-$\sigma$} AB limiting magnitudes in apertures with an area
that is three times as large as the seeing disk like in
\cite{2003ApJ...592..728S} where the imaging data used by
\cite{2005ApJ...619..697A} are described. We find that our images are
slightly shallower in all three bands used for the selection of LBGs
and thus, our larger correlation length may be due to incompleteness
at the faint end of the $23.3<R<25.3$ magnitude interval with the
\cite{2005ApJ...619..697A} LBG sample probing slightly deeper into the
luminosity function. The same problem certainly applies for the
$22.5<R<25.5$ and $22.5<R<26$ subsamples; this might be the reason for
the non-evolution of the correlation length for limiting magnitudes
$R_{\mathrm{lim}}>25$.

\subsection{Small-scale clustering}
\label{sec:halo}
\begin{figure}
  \resizebox{\hsize}{!}{\includegraphics{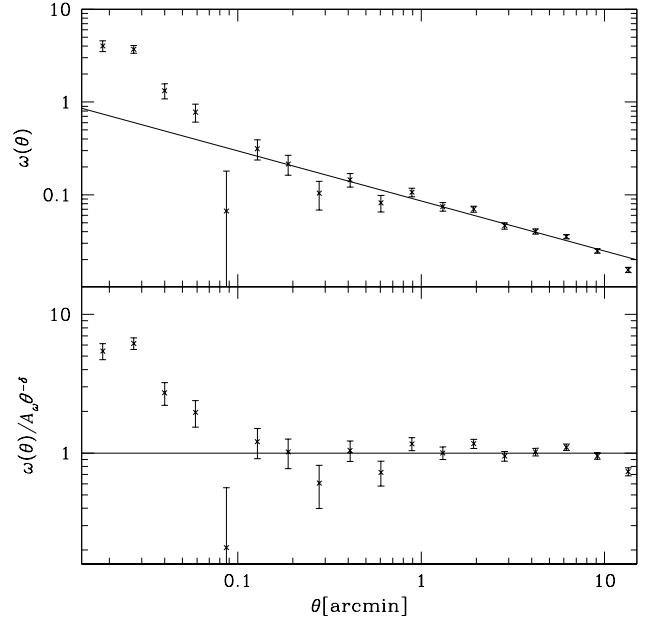}}
  \caption{\label{fig:excess}\emph{Upper} panel: Angular correlation
    function for $U$-dropouts with $22.5<R<26$. The solid line
    represents a power law fit to the data in the range
    $0\farcm1<\theta<10\arcmin$. \emph{Lower} panel: Ratios of the
    angular correlation function to the best-fit power law with a
    significant excess on small scales.}
\end{figure}

With the unprecedented statistical accuracy of our survey it is for
the first time possible to clearly detect an excess of the angular
correlation function of faint $U$-dropouts on small scales with
respect to a power law fit. In Fig.~\ref{fig:excess} the deviation of
the angular correlation function with respect to the power law fitted
at large to intermediate scales is shown. \cite{2005ApJ...635L.117O}
and \cite{2006ApJ...642...63L} report such a small-scale excess for
$z=4$ LBG samples from the Subaru/XMM-Newton Deep Field and the
GOODS fields, respectively.

This excess on small scales is interpreted in both studies as being
due to the contribution from a 1-halo term of galaxy pairs residing in
the same halos. We apply the halo model by \cite{2004MNRAS.347..813H}
to our data to have a direct comparison with the $z=4$ results from
\cite{2005ApJ...635L.117O} who use the same model.

In this model the angular correlation function of LBGs is
  calculated from the CDM angular correlation function by applying the
  following halo-occupation-distribution (HOD) for single galaxies:
\begin{equation}
  N_{\mathrm{g}}\left(M\right)=\begin{cases}
(M/M_1)^\alpha & \mathrm{for}\,\,M>M_{\mathrm{min}}\\
0 & \mathrm{for}\,\,M<M_{\mathrm{min}}
\end{cases}\,,
\end{equation}
and the following HOD for pairs of galaxies:
\begin{multline}
  \left<N_{\mathrm{g}}(N_{\mathrm{g}}-1)\right>(M)\\
=
\begin{cases}
N_{\mathrm{g}}^2\left(M\right) & \mathrm{if}\,\,N_{\mathrm{g}}\left(M\right)>1\\
N_{\mathrm{g}}^2\left(M\right)\log\left[4N_{\mathrm{g}}\left(M\right)\right]/\log4 & \mathrm{if}\,\,1>N_{\mathrm{g}}\left(M\right)>0.25\\
0 & \mathrm{otherwise}
\end{cases}\,,
\end{multline}
with $N_{\mathrm{g}}\left(M\right)$ being the number of galaxies in a
halo of mass $M$, $\left<N_{\mathrm{g}}(N_{\mathrm{g}}-1)\right>(M)$
being the number of galaxy pairs in a halo of mass $M$, and
$M_{\mathrm{min}}$, $M_1$, and $\alpha$ being the parameters of the
model. Furthermore, we calculate the number density of LBGs from this
model as described in \cite{2004MNRAS.347..813H}.

Applying a combined maximum likelihood fit to the angular correlation
functions and the number densities we find the best-fitting model
parameters for the different magnitude limited subsamples. From these
best-fit parameters we calculate the average mass of an LBG hosting
halo, $\left<M_{\mathrm{halo}}\right>$, and the average number of
galaxies inside this halo, $\left<N_{\mathrm{g}}\right>$ which are
also tabulated in Table~\ref{tab:clustering}.

Given the good agreement between our correlation functions at $z=3$
and the corresponding ones from \cite{2005ApJ...635L.117O} at $z=4$,
and given the structure growth of the dark matter density field
between $z=3$ and $z=4$ it is not surprising that we get slightly
larger halo masses. This would imply that star formation, which is
mostly responsible for the restframe UV flux, was slightly more
efficient at higher redshift. However, the evolution in halo mass is
rather small and not very significant. Judging from the residual
$\chi^2$ values for the best-fit parameters this model is still too
simple to account for the shape of the angular correlation functions
and the number densities simultaneously.

The mean number of LBGs per halo is well below one. This means that
there are a lot of halos which are not occupied by LBGs down to the
particular flux-limit. Nevertheless, this does not mean that these are
dark matter halos that do not host a galaxy. Massive galaxies that are
not actively forming stars may be very faint in the restframe UV and
have such red colours that they can easily escape our Lyman-break
selection technique. Other techniques incorporating near-IR data must
be used to select these populations \citep[see][]{2003ApJ...587L..79F,
  2003ApJ...587L..83V, 2004ApJ...617..746D}.

\section{Conclusions}
\label{sec:conclusions}
We measure the clustering properties of a large sample of $U$-dropouts
from the ESO Deep Public Survey with unprecedented statistical
accuracy at this redshift. 

Candidates are selected via the well-known Lyman-break technique and
the selection efficiency is investigated and optimised by means of
simulated colour catalogues. The angular correlation function of LBGs
is estimated over an area of two square degrees, depending on depth,
and a deprojection with the help of the photometric redshift
distribution yields estimates for the correlation lengths of different
subsamples.

We find clustering segregation with restframe UV-luminosity indicated
by a decreasing correlation length and a decreasing slope of the
correlation function with increasing limiting magnitude. The latter
result was reported at redshift $z=4$ and is now confirmed at redshift
$z=3$ for the first time. Furthermore, the unprecedented statistical
accuracy of our survey at $z=3$ allows us to study the small-scale
clustering signal in detail. We find an excess of the angular
correlation function on small angular scales similar to that found
previously at $z=4$.

Applying a halo model we find average masses for LBG-hosting halos at
$z=3$ which are slightly larger than literature values for $z=4$
implying decreasing star-formation efficiency with decreasing
redshift.

\begin{acknowledgements}
  We thank Dr. Leo Girardi for including the new WFI filters in the
  TRILEGAL code.

  Furthermore, we would like to thank Dr. Henry McCracken and Dr. Eric
  Gawiser for helpful discussions which improved this work a lot.

  This work was supported by the German Ministry for Education and
  Science (BMBF) through the DLR under the project 50 OR 0106, by the
  BMBF through DESY under the project 05 AV5PDA/3, and by the Deutsche
  Forschungsgemeinschaft (DFG) under the projects SCHN342/3-1 and
  ER327/2-1.

\end{acknowledgements}

\bibliographystyle{aa}

\bibliography{5880}

\end{document}